\documentclass[journal]{journal}
\usepackage{cite}
\usepackage{amsmath,amssymb,amsfonts}
\usepackage{algorithmic}
\usepackage{graphicx}
\usepackage{textcomp}
\usepackage{fancyhdr}
\usepackage{float}
\usepackage{multirow}
\usepackage{tabularx}
\usepackage{caption}
\usepackage{xcolor}
\usepackage{makecell} 
\usepackage{array}
\usepackage{booktabs}
\usepackage{bm}
\begin{document}
%
\title{StutterZero and StutterFormer: \\ End-to-End Speech Conversion for Stuttering Transcription and Correction}

\author{%
\uppercase{Qianheng Xu}\textsuperscript{1}%
\thanks{Manuscript received Month Day, Year; revised Month Day, Year. Corresponding author: Qianheng Xu (email: 26xuq@millburn.org).}%
\thanks{\textsuperscript{1}Millburn High School, Millburn, NJ 07041 USA.}
}

\markboth{Qianheng Xu \MakeLowercase{\textit{et al.}}: Preparation of Papers for IEEE TRANSACTIONS and JOURNALS}%
{}

\maketitle
\thispagestyle{empty}

\begin{abstract}

Over 70 million people worldwide experience stuttering, yet most automatic speech systems misinterpret disfluent utterances or fail to transcribe them accurately. Existing methods for stutter correction rely on handcrafted feature extraction or multi-stage automatic speech recognition (ASR) and text-to-speech (TTS) pipelines, which separate transcription from audio reconstruction and often amplify distortions.
This work introduces StutterZero and StutterFormer, the first end-to-end waveform-to-waveform models that directly convert stuttered speech into fluent speech while jointly predicting its transcription.
StutterZero employs a convolutional–bidirectional LSTM encoder–decoder with attention, whereas StutterFormer integrates a dual-stream Transformer with shared acoustic–linguistic representations. Both architectures are trained on paired stuttered–fluent data synthesized from the SEP-28K and LibriStutter corpora and evaluated on unseen speakers from the FluencyBank dataset.
Across all benchmarks, StutterZero had a 24\% decrease in Word Error Rate (WER) and a 31\% improvement in semantic similarity (BERTScore) compared to the leading Whisper-Medium model. StutterFormer achieved better results, with a 28\% decrease in WER and a 34\% improvement in BERTScore. 
The results validate the feasibility of direct end-to-end stutter-to-fluent speech conversion, offering new opportunities for inclusive human–computer interaction, speech therapy, and accessibility-oriented AI systems.
\end{abstract}

\begin{IEEEkeywords}
automatic speech recognition, deep learning, human-computer interaction, speech correction, speech processing, stuttering
\end{IEEEkeywords}
\IEEEpeerreviewmaketitle

\section{Introduction}
\label{sec:introduction}
Stuttering is a prevalent speech disorder that affects more than 70 million individuals worldwide \cite{ghai2021}. It manifests as interruptions in speech flow, including unintentional repetitions, prolongations, and pauses. Table \ref{tab:phonological_patterns} presents the five main phonological classifications of stuttering and examples of their symptoms \cite{kourkounakis2020a} \cite{basak2023}. Such interruptions can significantly hinder effective communication, frequently causing social anxiety, depression, and a diminished quality of life \cite{iverach2009}. In children, stuttering may lead to bullying and social isolation, exacerbating the emotional and psychological difficulties they encounter \cite{iverach2016}.

\begin{table}[!t]
	\centering
	\footnotesize
	\renewcommand{\arraystretch}{1.05}
	\begin{tabularx}{\columnwidth}{@{}l X@{}}
		\toprule
		\textbf{Classification} & \textbf{Example pattern} \\
		\midrule
		Sound repetition (SoundRep) & ``a-a-and'' \\
		Word repetition (WordRep)   & ``and and'' \\
		Interjection                & ``I think that---uhmm'' \\
		Block                       & ``I think\ldots\ \texttt{<pause>}\ \ldots that'' \\
		Prolongation                & ``sooooo'' \\
		\bottomrule
	\end{tabularx}
	\caption{Categories of stuttering with representative phonological examples.}
	\label{tab:phonological_patterns}
\end{table}

Fluent speech plays a critical role in daily communication, both in interpersonal situations and when engaging with intelligent voice-based technologies. Through a quality-of-life survey, people who stutter (PWS) showed a statistically significant decrease in emotional health and social function metrics \cite{kasbi2015}. Stuttering can also cause feelings of shame, fear, anxiety, and guilt \cite{bloodstein2021}. These challenges are exacerbated by the growing dependence on intelligent voice assistants, such as Google Home, Amazon Echo, and Apple Siri, which are designed to process fluent speech. As a result, people who stutter frequently encounter recognition errors, limited functionality, and exclusion when interacting with voice-controlled technologies.

For more severe forms of stuttering, automatic speech recognition (ASR) models, which transcribe speech, may insert unintended words or even truncate the speech due to a blocking stutter. Lea et al. \cite{lea2023} evaluated the speech of individuals with PWS using the Apple Speech framework, a production-level automatic speech recognition (ASR) system designed for fluent speakers. They reported a Word Error Rate (WER) of 19.8\%, indicating that 19.8\% of words in the ASR-generated transcript did not match the reference ground truth. They also reported a truncation rate of 23.8\%, where 23.8\% of utterances from these PWS were prematurely cut off \cite{lea2023}. Addressing this challenge, this study develops deep learning models that convert stuttered speech into fluent audio, going beyond transcription to reconstruct corrected acoustic signals. By generating fluent speech that preserves semantic content and improves intelligibility, these models aim to enhance communication for PWS and increase accessibility in human-machine interactions.

Previous work in the field of stutter correction can be separated into three categories: (1) Digital signal processing (DSP) and rule-based classifiers, (2) ASR and text-to-speech (TTS) pipelines, and (3) deep learning (DL) approaches.

\subsubsection{Conventional Digital Signal Processing Approaches}

DSP approaches utilize audio feature extraction methods to obtain condensed numerical features from complex audio signals. Then, a ruleset or a set of predetermined filters is applied to the features to determine which timeframes contain a stutter. Finally, these timeframes are cut out of the audio, removing the stutter. Some frequently utilized feature sets include Mel-Frequency Cepstral Coefficients (MFCCs), Linear Predictive Coding (LPC), and Linear Predictive Cepstral Coefficients (LPCCs) \cite{sheikh2022}. For instance, MFCC features are generated by first applying the Fast Fourier Transform (FFT) to convert an audio sample from the amplitude domain to the frequency domain, generating the power spectrum. After that, the Mel filter bank is used to map the power spectrum to Mel frequencies, employing a set of nonlinear triangular filters. This aligns the intensity of frequencies to match the nonlinearity of human auditory perception. Finally, a Discrete Cosine Transform (DCT) is used to generate the cepstral coefficients through the decorrelation of features. This pipeline is applied to every window of audio, usually 20–50 ms long \cite{sheikh2022} \cite{kn2020}.

K. N. et al. introduced a rule-based approach for stutter detection and removal by computing a “correlation factor” between adjacent audio windows using either MFCC or LPC features \cite{kn2020}. A high correlation value, empirically determined to be 0.92, was used to identify repeated or prolonged segments, prompting the deletion of redundant frames. The unusually high correlation factor was used to detect repeated audio patterns, such as recurring phonemes or extended silences. However, this approach was evaluated exclusively on repetition and prolongation stutters, without accounting for other common disfluencies such as blocks or interjections. Additionally, the experimental scope of the study was confined to only 60 repetition and 70 prolongation events obtained from a limited selection of audio files. As a result, its generalizability across diverse speakers, accents, or spontaneous conversational settings remains uncertain \cite{kn2020}. Table \ref{tab:classification_performance} summarizes the performance, demonstrating high within-sample accuracy.

\begin{table}[!t]
	\centering
	\scriptsize
	\setlength{\tabcolsep}{4pt}
	\renewcommand{\arraystretch}{1.1} 
	\begin{tabularx}{\columnwidth}{@{} l X c c c c @{}}
		\toprule
		\textbf{Stuttering Type} & \textbf{Features Used} & \textbf{Total} & \textbf{Removed} & \textbf{Retained} & \textbf{Accuracy} \\ 
		\midrule
		\multirow{2}{*}{Repetition} 
		& MFCC & 60 & 54 & 6 & 90.0\% \\ 
		& LPC  & 60 & 52 & 8 & 86.7\% \\ 
		\midrule
		\multirow{2}{*}{Prolongation} 
		& MFCC & 70 & 67 & 3 & 95.7\% \\ 
		& LPC  & 70 & 65 & 5 & 92.9\% \\ 
		\bottomrule
	\end{tabularx}
	\caption{Performance of DSP-based stutter removal methods using MFCC and LPC features across two stuttering types \cite{kn2020}.}
	\label{tab:classification_performance}
\end{table}

In some cases, low-level features such as MFCC, LPC, and LPCC are extracted from the audio and used as input to neural networks or machine learning models, which then classify whether a stutter has occurred at each point in time.

For example, StutterNet, introduced by Sheikh et al., is a multi-class Time Delay Neural Network (TDNN). Because TDNNs use windows of time-delayed inputs, they are especially well-suited to temporal dependencies such as MFCC speech features. While StutterNet was only able to detect stutters, it is plausible that similar systems could be used to flag sections of stuttered audio for removal \cite{sheikh2021}. Table \ref{tab:stutter_detection_summary} summarizes representative DSP-based systems, their methodological choices, and reported performance.

\begin{table*}[ht]
	\centering
	\footnotesize
	\setlength{\tabcolsep}{4pt} 
	\renewcommand{\arraystretch}{1.1} 
	\begin{tabularx}{\textwidth}{X X X X X X}
		\toprule
		\textbf{Author(s)} & \textbf{Dataset} & \textbf{Feature Extraction} & \textbf{Detection Method} & \textbf{Stutter Type(s)} & \textbf{Best Accuracy} \\
		\midrule
		Mishra et al., 2021 & UCLASS & MFCC, RMSE & Deep neural network & Repetition, Prolongation, Block & 86.67\% \\
		Dash et al., 2018 & Private human speech samples & Amplitude & Deep neural network & Prolongation & 86\% \\
		Shonibare et al., 2022 & Private human speech samples & log-Mel Spectrograms & CNN & Repetition, Block, Prolongation & 71.24\% reduction in WER \\
		\bottomrule
	\end{tabularx}
\caption{Summary of studies on stutter detection, including datasets, feature extraction methods, detection approaches, stutter types analyzed, and best reported accuracies~\cite{kn2020}.}
	\label{tab:stutter_detection_summary}
\end{table*}

\subsubsection{ASR \& TTS-based Approaches}
\begin{figure*}[!t]
	\centering
	\includegraphics[width=\textwidth]{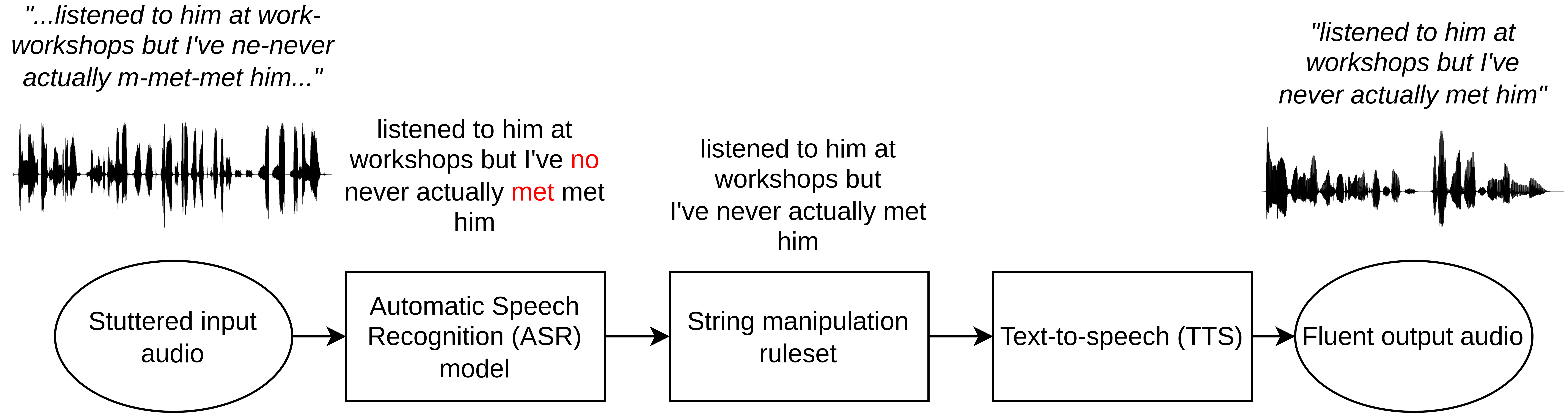}
	\caption{General pipeline of an ASR-TTS approach.}
	\label{fig:asrintro}
\end{figure*}

The second approach to stutter correction involves fine-tuning or training an ASR model on stuttered speech such that the ASR model can generate transcripts of that speech. The ASR model can either explicitly transcribe the stutter into text or ignore the stuttered portions, producing a fluent transcript. That transcript can be filtered with rule-based systems and finally passed through a TTS model to produce fluent audio sequences. Figure \ref{fig:asrintro} shows the general pipeline to achieve ASR \& TTS-based stutter correction. This approach helps to omit artifacts caused by splicing audio in the DSP-based approaches, since TTS models are generating new audio sequences. However, DSP methods can preserve speaker prosody more naturally, as they simply edit the original speech. For a TTS model to mimic the tone and prosody of the original speaker, it would need more utterances to fine-tune on.

Mujtaba \& Mahapatra describe fine-tuning Whisper-small, a state-of-the-art ASR model pretrained on fluent speech corpora. To allow for efficient training, low-rank adaptation (LoRA) was used as a form of parameter-efficient fine-tuning. Whisper-small was fine-tuned on ground truth transcripts provided by the FluencyBank and private HeardAI datasets. They reported that Whisper-Small achieved a WER of 33.88\% without any training or fine-tuning. After fine-tuning Whisper-Small using the aforementioned LoRA methods, it achieved a WER of 9.39\% \cite{mujtaba2025}.

Due to the adaptability of fine-tuned ASR models, there has been research into fine-tuning pre-trained ASR models for other speech impairments such as dysarthria. Wang et al. fine-tuned wav2vec2.0 and HuBERT ASR models on dysarthria datasets. They achieved the best WER of 16.53\% by fine-tuning a pre-trained wav2vec2.0 ASR model on a dysarthria corpus augmented by a generative adversarial network. Even without any data augmentation, they obtained a WER of 22.25\% on a fine-tuned wav2vec2.0 model and a WER of 21.10\% on a fine-tuned HuBERT model \cite{wang2024a}.

\subsubsection{Deep Learning \& End-to-end Approaches}
Deep learning approaches for speech recognition and speech conversion have been the subject of extensive investigation. Speech conversion focuses on modifying or transforming a speech signal, such as changing the speaker's voice or style, while preserving the original linguistic content. In contrast, speech recognition involves accurately transcribing spoken language into text \cite{triantafyllopoulos2023} \cite{walczyna2023} \cite{wang2019}. Deep learning approaches for speech processing typically involve training neural networks or other computational models, eliminating the need for manual feature engineering. Recent advancements in deep learning have popularized end-to-end models, systems that learn to perform the entire correction process directly from input data without requiring handcrafted features or intermediate steps.

While there is considerable work on end-to-end speech conversion of fluent speech and other speech impairments such as dysarthria, there is little research on deep learning or end-to-end stutter correction. This paper first reviews deep learning and end-to-end methods for fluent speech recognition, conversion, and the correction of dysarthric speech.

The popularity of end-to-end models stems from their ability to consolidate the entire training and inference pipeline into a single model optimized with one objective function that directly reflects the training goal. In contrast, a traditional DSP-based pipeline might first extract MFCC features from audio and then train a neural network to classify whether each frame contains a stutter. The neural network in this case is trained for frame-level accuracy in detecting stutters, which does not align with removing stutters at the speech sequence level \cite{chan2016} \cite{graves2014}. The lack of manual feature engineering in end-to-end models also allows for greater generalizability and adaptability to similar problems \cite{hannun2014}. Finally, end-to-end models such as encoder-decoder and RNN-Transducer models are suitable for sequence-to-sequence tasks such as speech conversion, as they do not require alignment of acoustic sequences to ground truth transcripts and are very flexible with input and output sequence lengths \cite{wang2019}.

For example, Toshniwal et al. developed an attention-based encoder-decoder model inspired by recurrent neural networks \cite{toshniwal2017}. A speech encoder is built on a deep bidirectional long short-term memory (BiLSTM) network, which produces a sequence of abstract hidden representations. These hidden representations are passed into the character decoder, which is a single-layer LSTM that predicts the most likely letter uttered \cite{toshniwal2017}.

Wang \cite{wang2022} introduced an end-to-end encoder-decoder for dysarthric speech correction. Three multitask encoders were used: a content encoder to learn underlying semantic meaning, a prosody encoder to learn and correct audio features salient to dysarthria, and a speaker encoder to capture prosody and recreate the tone and voice of the original speaker. A decoder aggregates hidden representations from all three layers and generates acoustic features from which audio can be reconstructed using a vocoder \cite{wang2022}.

Despite these advances, work on stuttering remains limited. Most deep learning research to date has focused on event-level detection and classification of disfluencies rather than the synthesis of corrected audio. Approaches that combine automatic speech recognition with text-to-speech generation can yield fluent renditions, but they are constrained by a lack of transcription accuracy or generalization across all categories of stuttering. Direct acoustic-to-acoustic correction of stuttered speech into fluent speech remains an underexplored problem. This gap motivates the present study, which introduces two end-to-end models, \emph{StutterZero} and \emph{StutterFormer}, that jointly address disfluency detection, transcription, and fluency restoration. By explicitly targeting corrected audio generation while preserving semantic content, these models extend prior deep learning approaches toward real-time stutter correction and inclusive speech technology.

\section{Methodology}
This section presents the methodological framework of the study, which is organized into four main parts. First, the datasets used for training and evaluation are described, along with the corresponding data preprocessing and cleaning procedures. Second, the ASR–TTS baseline pipeline is outlined, in which a pretrained ASR system is adapted to stuttered speech and fluent output is resynthesized using a TTS model. Third, two proposed end-to-end models, StutterZero and StutterFormer, are introduced; these models directly transform disfluent speech into fluent speech through multitask encoder–decoder architectures. Finally, the training configuration, optimization procedures, and cross-validation strategy employed to ensure robustness and reproducibility are detailed.

\subsection{Data}

This research uses two datasets for training: Stuttering Events in Podcasts (SEP-28K) and LibriStutter. The SEP-28K dataset contains 23 hours of naturally occurring stuttering events, divided into 28,000 clips, each lasting 3 seconds \cite{lea2023}. While this dataset does not contain any ground truth for what the fluent speech should be, it includes labels classifying every stutter into the categories shown in Table \ref{tab:phonological_patterns}. All audio recordings were taken from public podcasts with people who stutter at a standard sampling rate of 16 kHz. After all cleaning and processing, about 14 hours of raw audio data were left.

About 20 hours of artificially produced stuttered audio are included in the LibriStutter dataset. This corpus was created by splicing, cutting, duplicating, and performing other manipulations on the fluent LibriSpeech corpus to mimic the prosodic characteristics of stuttering \cite{kourkounakis2020a}. SEP-28K does not include fluent reference transcripts, whereas LibriStutter contains paired fluent transcriptions that facilitate supervised training.

\subsubsection{Data Cleaning and Preprocessing}
To ensure that all audio contained speech, audio files labeled "Music" or "NoSpeech" were removed from the SEP-28K dataset. To account for bias or skew in the frequency of each type of stutter, data resampling was conducted on SEP-28K to balance out stutter categories that may have been less common.

After cleaning, approximately 34 hours of stuttered audio were retained across the SEP-28K and LibriStutter datasets, with no overlap in speakers or transcripts between the two corpora.

The University College London Archive of Stuttered Speech (UCLASS) and the FluencyBank dataset are the two other well-known stuttering datasets \cite{zotero-item-1105} \cite{bernsteinratner2018}. However, both datasets have some limitations. The UCLASS dataset does not provide fluent "ground truth" transcripts for the stuttered speech, so there are no reference transcripts to fine-tune Whisper-Small on.

The FluencyBank dataset contains the Voices-AdultsWhoStutter (Voices-AWS) corpus, which was de-identified and made publicly available to researchers who created a free account. StutterZero and StutterFormer were later tested on the Voices-AWS subset of the FluencyBank dataset for validation purposes.

\subsubsection{Data Splitting and Cross Validation for Training}
To train and validate all three approaches (ASR-TTS, StutterZero, and StutterFormer) introduced in this paper, the combined dataset of audio-transcript pairs from SEP-28K and LibriStutter was randomly sampled and split into training (80\%), test (10\%), and validation (10\%) sets. The training and testing splits were used in a five-fold cross-validation setup, with the validation set being used for a fair comparison between all three approaches after training. All training for this study was conducted on an NVIDIA RTX 3080 with 10 gigabytes of VRAM.

\subsection{ASR-based approach}
To obtain fluent speech data to train StutterZero and StutterFormer as end-to-end models, this research also developed an auxiliary pipeline as described below to first generate fluent speech data, effectively "completing" both datasets:

\begin{enumerate}
	\item Stuttered audio and fluent transcripts from LibriStutter were used to fine-tune a Whisper-Small ASR model that was originally trained on fluent speech.
	\item The fine-tuned Whisper-Small model was applied to the SEP-28K dataset, generating fluent transcripts for all SEP-28K audio clips.
	\item A pretrained MeloTTS text-to-speech model was applied to all fluent transcripts from both datasets, generating accurate and clear fluent audio sample counterparts.
\end{enumerate}

\subsection{Whisper-Small Architecture}

This study fine-tuned Whisper-Small, a lightweight derivative of the Whisper family of ASR models. Audio data from both datasets are in the amplitude domain, though Whisper-Small accepts a spectrogram in the frequency domain as input \cite{radford2022}.

\begin{figure}[!t]
	\centering
	\includegraphics[width=\columnwidth]{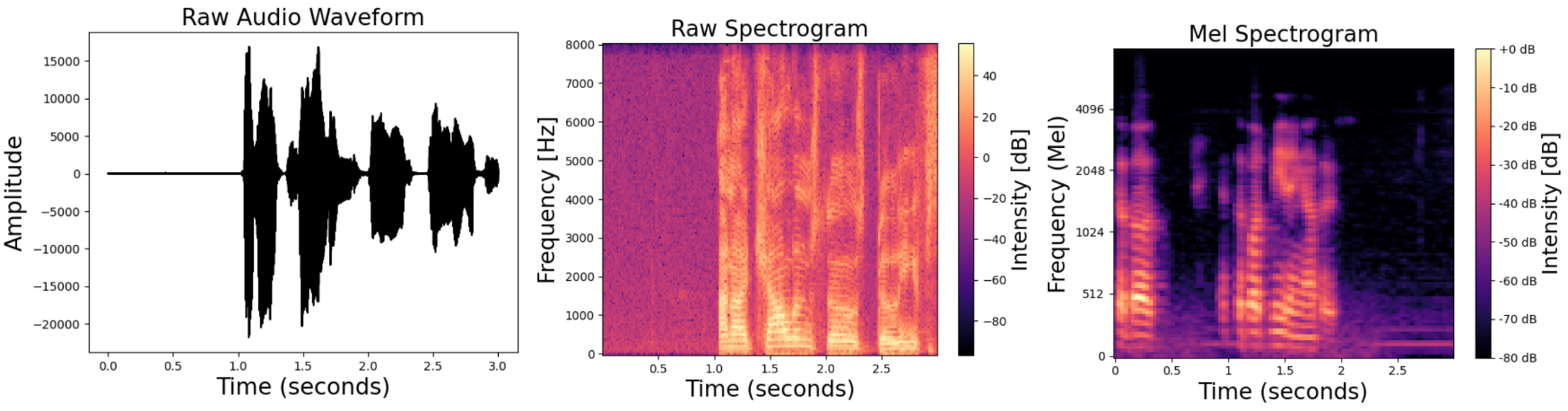}
	\caption{Conversion from waveform to log-Mel spectrogram. The waveform (left) shows amplitude over time. After STFT, the spectrogram (middle) represents frequency over time. Applying the Mel filterbank produces a log-Mel spectrogram (right), aligned with human auditory perception.}
	\label{fig:mel}
\end{figure}

Whisper-Small’s WhisperFeatureExtractor uses a Short-Time Fourier Transform (STFT), computing the Fast Fourier Transform (FFT) on overlapping segments, or “windows,” of the audio as it slides across the entire signal. Then, the spectrogram is converted into a log-Mel spectrogram by mapping the frequencies from the “vanilla” spectrogram onto the Mel scale. This is done because human hearing does not perceive pitch in a linear manner; humans are more attuned to changes in lower pitches than in higher pitches. The linear frequency spectrogram is passed through a Mel filter bank, which applies a series of triangular filters to aggregate spectral energy within perceptually relevant frequency bands. Once a spectrogram is transformed into a log-Mel spectrogram, equal intervals on the Mel scale reflect equal perceived differences in pitch. The practical effect is that Mel spectrograms highlight the frequencies of human speech while minimizing the intensity of background noise, allowing the model to concentrate solely on the speech signals \cite{stevens1937}. Figure \ref{fig:mel} displays visualized spectrograms of the conversion process used to obtain a log-Mel spectrogram.

Whisper-Small computes log-Mel spectrograms with 25-ms-long windows that move forward by 10 milliseconds at every time step.

Whisper-Small is a Transformer encoder-decoder model featuring 12 layers in both the encoder and decoder, a hidden width of 768 units, and 12 attention heads, amounting to around 244 million parameters in total. The encoder layers consist of a self-attention mechanism and fully connected hidden layers. The encoder generates a context vector, which is passed via cross-attention to each of the 12 decoder attention blocks. Each decoder layer contains self-attention, cross-attention, and a fully connected network \cite{radford2022}. At every time step, the autoregressive decoder is fed the aggregated output tokens from all previous time steps so that it can predict the most likely following token.

Whisper’s built-in byte-pair encoding (BPE) tokenizer converts the numeric predictions of Whisper, called tokens, back into readable text.

\subsection{Whisper-Small Fine-tuning}
To fine-tune Whisper-Small, this study aggregated all stuttered audio samples and normalized the sampling rate to 16 kHz. For LibriStutter audio clips, longer audio samples are truncated after 30 seconds.

Fine-tuning begins with Whisper-Small’s pretrained weights to take advantage of the accurate general transcription properties that Whisper is already trained for. The evaluation metric is Word Error Rate (WER), which is formally defined in Equation \eqref{eq:wer} \cite{hunt1990}.

\begin{equation}
	\text{WER} = \frac{\text{substitutions} + \text{deletions} + \text{insertions}}{\text{number of words in the reference}}
	\label{eq:wer}
\end{equation}

$S$ is the number of substitutions (words in the reference that are replaced with incorrect words in the hypothesis), $D$ is the number of deletions (words in the ground truth that are missing from the predicted speech), and $I$ is the number of insertions (extra words in the predicted speech that are not in the ground truth).

Training runs for a maximum of 10,000 steps with a learning rate of 1e-5 and a batch size of 8. Gradient accumulation over two steps is used to simulate a larger batch size. Gradient checkpointing and mixed precision are enabled to save memory and speed up training. Evaluation occurs every 1,000 steps, and the model weights with the lowest WER are saved.

	\subsection{StutterZero Model Architecture}

End-to-end models have seen significant usage in other speech-based tasks such as translation and transcription. These models are characterized by directly converting an input signal to an output signal without any intermediate feature engineering or representation.

This study introduces an autoregressive, end-to-end, multitask model named StutterZero. Inputs consist of 80-channel log-Mel spectrograms computed with a 50 ms window length and a 12.5 ms frame shift. The encoder converts the input log-Mel spectrogram into a higher-level representation called a context vector. The context vector is a numerical representation of features that the trained encoder determines to be relevant. While typical encoder-decoder models use one encoder and one decoder, this study proposes a multitask decoder in which two decoders are forced to predict different data types. During training, a multitask model minimizes a joint loss function, forcing the decoders to learn more generalized patterns that benefit all constituent losses \cite{zhang2021}. The spectrogram decoder predicts the fluent spectrogram signal, while the transcript decoder predicts the grapheme being uttered.

\begin{figure*}[t]
	\centering
	\includegraphics[width=\textwidth]{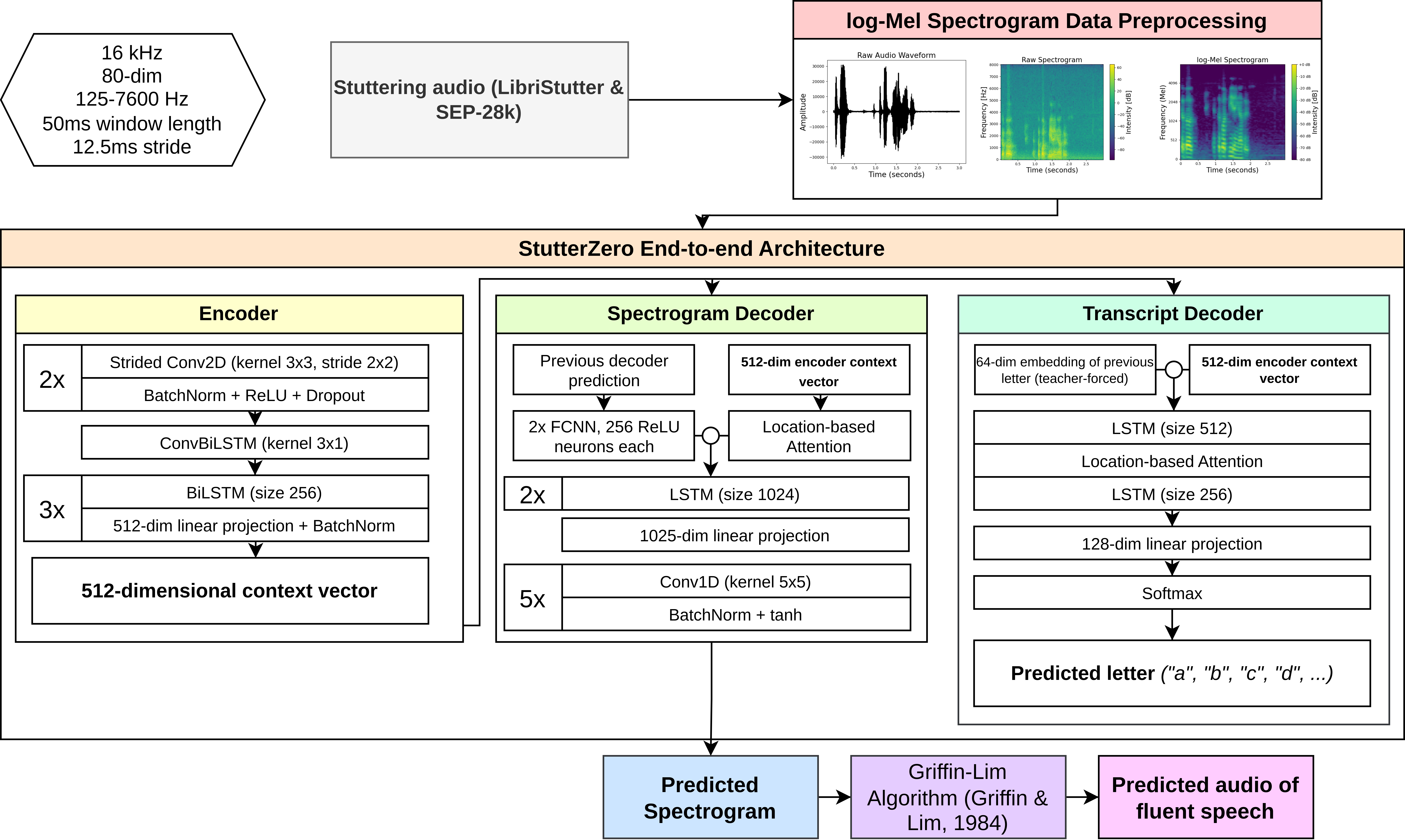}
	\caption{Overview of the StutterZero end-to-end architecture. A convolutional BiLSTM encoder compresses these features into a 512-dimensional context vector, which is shared between two multitask decoders. The spectrogram decoder predicts fluent spectrogram frames that are reconstructed into audio using the Griffin–Lim algorithm, while the transcript decoder predicts grapheme tokens.}
	\label{fig:stutterzero_architecture}
\end{figure*}

The spectrogram decoder generates the spectrogram one frame at a time, using the context vector along with the previously predicted frames as contextual input to append each new predicted frame to the output spectrogram. Similarly, the transcript decoder uses previously predicted tokens to predict the following grapheme tokens. Figure \ref{fig:stutterzero_architecture} displays an overview of the model architecture.

\subsubsection{StutterZero Encoder}
The encoder consists of two convolutional blocks, each including a two-dimensional convolution layer with a 3×3 kernel, a 2×2 stride, and batch normalization. To effectively capture both spatial and temporal features of the input sequence, a convolution-augmented bidirectional long short-term memory network (ConvBiLSTM) is employed. Log-Mel spectrograms provide information about both the frequency content of a signal and the timing of its occurrence. Unlike traditional LSTMs, the ConvBiLSTM replaces standard matrix multiplication with a convolution operation, which promotes the learning of local spatial patterns in two-dimensional data. The network includes mechanisms analogous to the standard LSTM gates—input, forget, and output gates—as well as candidate cell states and hidden states. Each gate and state use convolutional kernels and trainable biases to process the current input in combination with the hidden state from the previous time step, enabling the model to capture complex spatiotemporal dependencies \cite{shi2015}.

The encoder generates a 512-dimensional context vector, which is used as input for both decoders.

\subsubsection{StutterZero Spectrogram Decoder}
The spectrogram decoder uses the previously predicted spectrogram and the current context vector as inputs. Two fully connected layers form a pre-net that compress and transform the previous spectrogram frames into a lower-dimensional representation. A well-trained pre-net simplifies the input and extracts the most salient features. If the raw spectrogram frames were used, the decoder might "shortcut" the learning process by copying the previous frame or minimally modifying it \cite{shen2018} \cite{wang2017}.

StutterZero employs a location-based attention mechanism, which is an extension of classical content-based attention. In content-based attention, the model computes an attention score between a query (representing the decoder’s current state) and a key (representing the encoder's output at any time step). The score measures how relevant each encoder state is to the current decoding step. The attention weights are then obtained by normalizing these scores using a softmax function. These weights form a probability distribution that determines how much "attention" the decoder should pay to each encoder state when predicting the next output. Finally, a weighted context vector is used for output prediction \cite{chorowski2015}.

Because speech data are only read in one direction (monotonically), classical content-based attention may "jump around" erratically without respecting the flow of time. Location-based attention also computes a set of features using concatenated previously calculated attention weights via a one-dimensional convolution.

The location-based features are included in the attention score along with another set of trainable weights. In this way, the decoder is informed by past attention behavior and adjusts its focus accordingly to maintain a monotonic flow of data.

Once the spectrogram decoder predicts a fluent spectrogram signal, the Griffin–Lim algorithm is used to reconstruct the phase data and generate an audio signal in the amplitude domain \cite{griffin1984}.
	
\subsubsection{StutterZero Transcript Decoder}	
Previous work in text-to-speech and fluent speech conversion models has demonstrated the efficacy of multitask decoder architectures. Multitask training sums the loss for both the spectrogram and transcript decoders, creating a joint loss function that forces the training process to optimize the loss on both decoders \cite{radford2022} \cite{toshniwal2017} \cite{zhang2021}.

In the case of the transcript decoder, adding an objective function at the grapheme level may allow StutterZero to learn more intricate orthographic features of a word to correctly distinguish between allophones and homophones.

The transcript decoder also uses a teacher-forced embedding of the previous text as its input during training. Instead of using its previous prediction as input for generating the next token, the true ground-truth token from the training data is fed as input to the next step. This stabilizes and speeds up training because the model always conditions on the correct previous tokens. Additionally, it avoids extreme divergence and error in the early stages of training, when incorrect predictions may accumulate.

\subsubsection{StutterZero Training}
The spectrogram decoder uses mean squared error (MSE) loss, where the model minimizes the difference between the ground-truth spectrogram and the predicted fluent spectrogram across all frequency bins and time steps \cite{rafaely2025, wang2009}.

Cross-entropy loss is used for the transcript decoder because grapheme prediction is a categorical task. This loss measures how well the model predicts the correct token given all previous tokens and the input log-Mel spectrogram.

To combine the loss functions for both decoders, a bespoke loss function (described in Equation \ref{eq:combinedloss}) is defined by summing two components: the mean-squared error (MSE) loss between the fluent and stuttered spectrogram frequency bins, and the cross-entropy loss between the predicted token probability distribution and the ground-truth tokens.
\setlength{\abovedisplayskip}{0pt}
\setlength{\belowdisplayskip}{0pt}
\begin{equation}
	\textit{L}_{\text{total}} = \textit{L}_{\text{MSE}} + \textit{L}_{\text{CE}}
	\label{eq:combinedloss}
\end{equation}
	StutterZero used standard stochastic gradient descent with the Adam optimizer for training, starting with a learning rate of $1e^{-4}$ and a weight decay of $1e^{-6}$. The weight decay discourages large weights by incorporating the L2 norm of the weights into the loss function. Table \ref{tab:szparams} shows all the hyperparameters used in the Adam optimizer.

	\begin{table}[H]
		\centering
		\footnotesize
		\setlength{\tabcolsep}{1.5pt} 
		\renewcommand{\arraystretch}{1.3} 
		\begin{tabular}{l c c c c}
			\toprule
			\textbf{Parameter} & \textbf{Value} \\
			\midrule
			Learning Rate & $1e^{-4}$ &  \\
			Weight Decay & $1e^{-6}$ & \\
			Batch Size & $3$ \\
			Betas (Momentum parameters) & $(0.9, 0.999)$ \\
			Epsilon & $1e^{-6}$ \\
			\bottomrule
		\end{tabular}
		\caption{Training hyperparameters used for the StutterZero Adam optimizer.}
		\label{tab:szparams}
	\end{table}
	
			\subsection{StutterFormer Model Architecture}
	
\begin{figure*}[h]
	\centering
	\includegraphics[width=\textwidth]{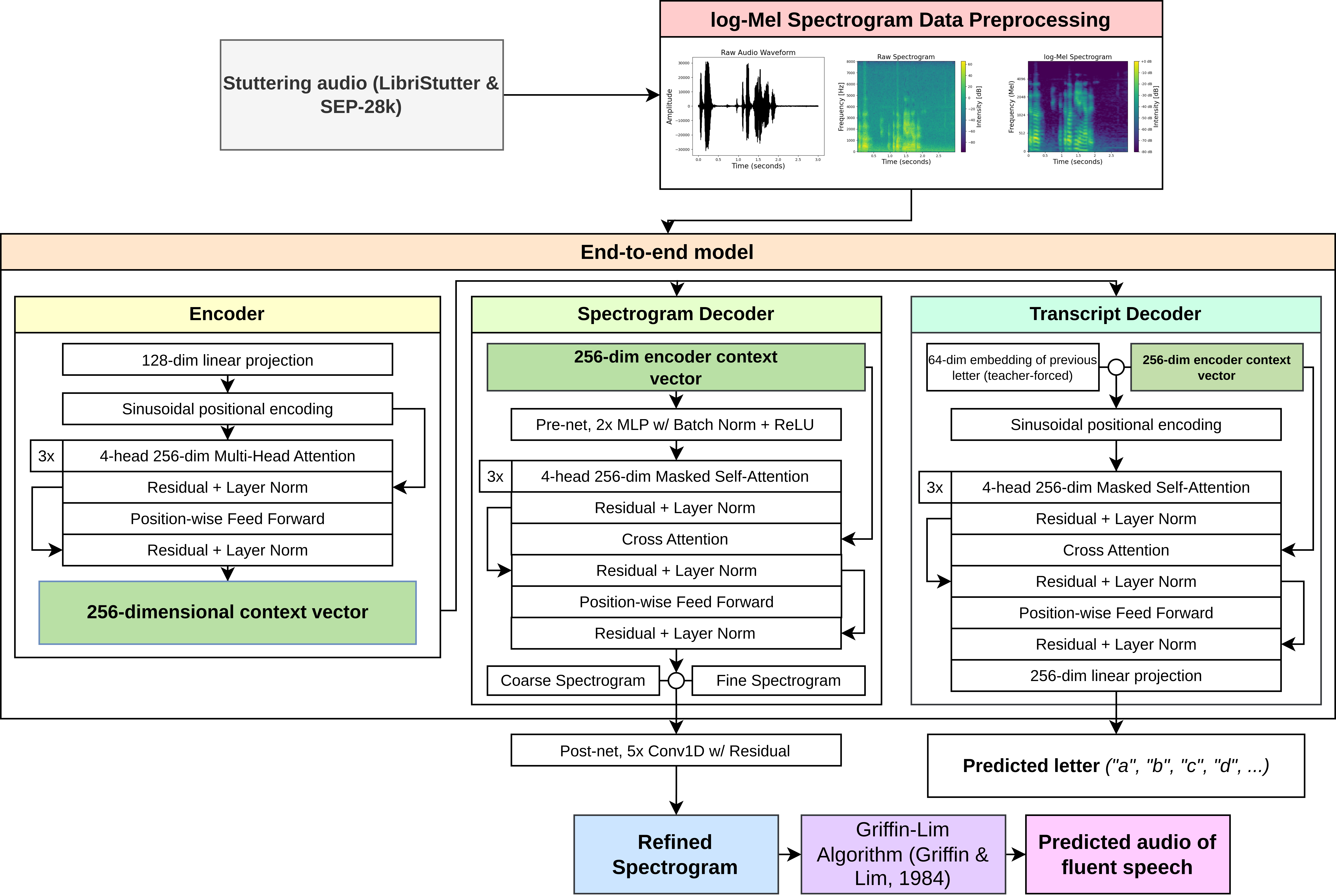}
	\caption{Overview of the StutterFormer end-to-end model. The encoder projects features into a 128-dimensional space, applies sinusoidal positional encoding, and processes them through stacked multi-head attention and feed-forward layers to produce a 256-dimensional context vector. This vector is shared between two multitask decoders. The spectrogram decoder refines fluent spectrograms via masked self-attention, cross attention, and convolutional post-nets, while the transcript decoder predicts grapheme tokens using masked self-attention and cross attention layers. The Griffin–Lim algorithm reconstructs waveforms from refined spectrograms, jointly optimizing fluency restoration and transcription.}
	\label{fig:stutterformer_architecture}
\end{figure*}

	StutterFormer maintains the same multitask architecture as StutterZero, but switches out internal layers for the Attention mechanism found in Transformers as described in \cite{vaswani2023}. Figure \ref{fig:stutterformer_architecture} displays a high-level summary of StutterFormer's architecture.

\subsubsection{Multi-Head Attention}
Multi-head attention serves as the fundamental building blocks of StutterFormer. Compared to single-head attention, which computes attention in a single attention distribution, multi-head attention projects queries, keys, and values into multiple subspaces, applies attention in parallel, and then recombines the results. This makes it possible for the model to take a joint attention function to data from several learned subspaces. For instance, one head may focus on short-range syntactic dependencies only while another head tracks long-range semantic connections. This increase in flexibility allows multi-head attention to learn a greater breadth of information while keeping the parameter count relatively equal to a larger single-head attention module \cite{vaswani2023}. 

\subsubsection{StutterFormer Encoder}
The encoder input remains a log-Mel spectrogram. Because a Transformer processes all input frames in parallel, it does not have a sense of order by default. Sinusoidal positional encoding gives each time step a deterministic vector (based on the position index) for the embedding at each time step. Residual and layer normalization layers are used between the multi-head attention and feed-forward layers to mitigate vanishing or exploding gradients \cite{borawar2023} \cite{he2015}. Three multi-head attention units are utilized in the encoder, each consisting of four heads. Similarly to StutterZero, the encoder outputs a context vector -- a learned hidden representation of the input.

\subsubsection{StutterFormer Decoders}
The spectrogram decoder uses the context vector from the input and uses a similar pre-net architecture as StutterZero to learn a lower-dimension representation. Masked multi-head self-attention is used to prevent the decoder from "looking ahead" at future frames that have not been predicted yet. After applying the mask, the decoder can only attend to itself and past frames. Cross attention utilizes queries from the preceding decoder layer, while the keys and values are derived from the original encoder context vector. This allows the decoder to observe the entire encoder context when deciding the next frame.

Finally, a post-net consisting of five 1-dimensional convolutions refines the predicted mel spectrogram to denoise and sharpen the final audio signal \cite{shen2018} \cite{ren2019}. 

The transcript decoder uses a very similar architecture to the spectrogram decoder and employs the same Transformer decoder architecture.

\subsubsection{StutterFormer Training}
Like StutterZero, StutterFormer employs a hybrid loss function that combines multiple weighted losses.
The spectrogram decoder also uses MSE loss between the predicted and ground truth spectrograms. In addition to MSE loss, the spectrogram decoder also computes a mean absolute error (MAE) loss. MAE loss measures the absolute difference of the predicted values and target values without squaring. When tested with spectrogram applications, MAE loss promotes sharper spectrograms that prevent the predicted fluent speech from sounding slurred \cite{guso2022}. The transcript decoder also uses cross-entropy loss.

StutterFormer is trained on a cosine annealing with warm restarts scheduler. This approach periodically "restarts" the learning rate to help escape local minima and explore the loss landscape more effectively \cite{liu2022} \cite{jiang2024} \cite{cazenave2022a}. Like StutterZero, StutterFormer was trained for 1000 epochs.
	\begin{table}[H]
	\centering
	\footnotesize
	\setlength{\tabcolsep}{1.5pt} 
	\renewcommand{\arraystretch}{1.3} 
	\begin{tabular}{l c c c c}
		\toprule
		\textbf{Parameter} & \textbf{Value} \\
		\midrule
		Learning Rate & $1e^{-4}$ &  \\
		Weight Decay & $1e^{-5}$ & \\
		Batch Size & $3$ \\
		Betas (Momentum parameters) & $(0.9, 0.98)$ \\
		Epsilon & $1e^{-6}$ \\
		$T_0$ (Initial restart epochs) & 50\\
		$T_{mult}$ (Period multiplication factor) & 2\\
		\bottomrule
	\end{tabular}
	\caption{Training hyperparameters used for the cosine annealing scheduler and optimizer.}
	\label{tab:sfparams}
\end{table}

\section{Results}
\subsection{Benchmarking}
In addition to the ASR-based and end-to-end approaches used in this study, state-of-the-art fluent-speech ASR models were also evaluated to establish a baseline for how accurately current speech recognition systems transcribe stuttered speech. This study used the Whisper-Tiny, Whisper-Small, and Whisper-Medium pretrained fluent ASR models as baselines \cite{radford2022}. Any larger models such as Whisper-Large could not be tested due to memory and hardware limitations. The 10\% validation data split was used to assess each of the six models.

Three metrics were used to evaluate all models: Word Error Rate (WER), Character Error Rate (CER), and BERTScore. CER is defined in Equation \eqref{eq:cer} as the proportion of incorrectly predicted characters compared to the ground truth string. To assess the semantic similarity between the ground truth and predicted utterances, a BERTScore is calculated using pre-trained contextual embeddings from the Bidirectional Encoder Representations from Transformers (BERT) language model \cite{zhang2020}. Cosine similarity is used as a metric to quantify similarity between high-dimensional embedding vectors. The BERTScore defines how similar the two strings are semantically, meaning it is more forgiving towards minor transcription mistakes that still preserve the overall meaning of the utterance \cite{zhang2020}.

\begin{equation}
	\text{CER} = \frac{\text{char. substitutions} + \text{char. deletions} + \text{char. insertions}}{\text{number of characters in the reference}}
	\label{eq:cer}
\end{equation}	
	
\begin{figure}[h]
	\centering
	\includegraphics[width=0.48\textwidth]{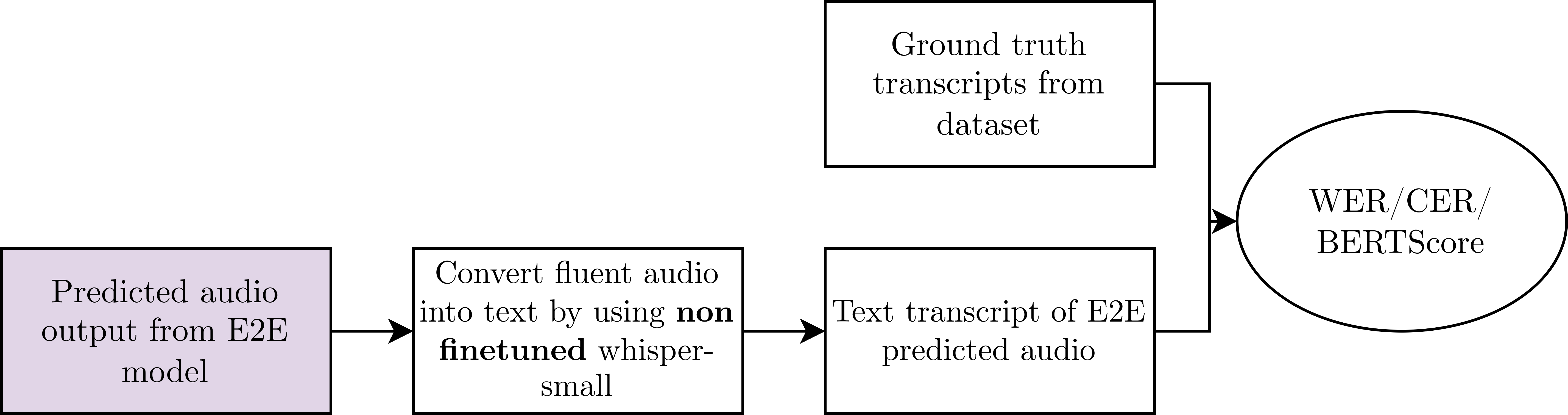}
	\caption{The validation pipeline. An untouched Whisper-Small ASR model acted as "judge" to evaluate speech clarity and intelligibility.}
	\label{fig:valid}
\end{figure}

Because both approaches in this research produce audio signals, the audio signals must be converted back to text before any word or character-level evaluation. This study used an unmodified, pretrained copy of the Whisper-Small ASR model to transcribe the predicted fluent sequences from both the ASR and end-to-end approaches, as shown in Figure \ref{fig:valid}. This was an attempt to simulate what an untrained, average English-speaking person would interpret the fluent speech to be. The predicted transcripts of the fluent speech were compared with ground truth transcripts to calculate the final metrics. Table \ref{tab:metrics} displays the mean WER, CER, BERTScore Precision, and their standard deviations in the validation data split.

\begin{table*}[ht]
	\centering
	\footnotesize
	\setlength{\tabcolsep}{6pt} 
	\renewcommand{\arraystretch}{1.3} 
	\begin{tabularx}{\textwidth}{X c c c}
		\toprule
		\textbf{Model} & \textbf{Mean WER} & \textbf{Mean CER} & \textbf{Mean BERTScore Precision} \\
		\midrule
		Whisper-Tiny & $0.415 \pm 0.049$ & $0.227 \pm 0.033$ & $0.5768 \pm 0.027$ \\
Whisper-Small & $0.370 \pm 0.037$ & $0.171 \pm 0.027$ & $0.5956 \pm 0.017$ \\
Whisper-Medium & $0.361 \pm 0.032$ & $0.162 \pm 0.022$ & $0.601 \pm 0.017$ \\
ASR-based & $0.04 \pm 0.01$ & $0.02 \pm 0.01$ & $0.9516 \pm 0.04$ \\
StutterZero (end-to-end) & $0.116 \pm 0.013$ & $0.110 \pm 0.052$ & $0.9174 \pm 0.034$ \\
StutterFormer (end-to-end) & $0.08 \pm 0.03$ & $0.07 \pm 0.011$ & $0.9411 \pm 0.012$ \\
		\bottomrule
	\end{tabularx}
	\caption{Combined WER, CER, and Mean BERTScore Precision metrics for all models.}
	\label{tab:metrics}
\end{table*}

Table \ref{tab:metrics} shows that all three proposed models—ASR-based, StutterZero, and StutterFormer—outperform the best Whisper baseline (Whisper-Medium) across both WER and CER metrics. The Whisper-Medium model achieves a WER of 0.361 and a CER of 0.162, while the ASR-based approach dramatically reduces these errors to 0.04 and 0.02, respectively. This indicates a substantial improvement in transcription accuracy and character-level precision.

\subsection{Statistical Significance}
The non-parametric, two-sided Wilcoxon Signed-Rank Test is used to assess the statistical significance of the improvements achieved by this research compared to the baseline Whisper performance. This test is chosen due to its widespread use and established precedent in evaluating significant differences between models \cite{rainio2024}. Comparing the performance of the ASR-based approach with Whisper-Medium (the best performing Whisper model), the Wilcoxson Test returns a test statistic of $77631.0$, a p-value of $<1\mathrm{e}^{-100}$, significant at $\alpha = 0.05$. Running the test comparing the performance of the end-to-end StutterZero and StutterFormer models against Whisper-Medium also returns a p-value of $<1\mathrm{e}^{-100}$, significant at $\alpha = 0.05$. This shows that both StutterZero and StutterFormer demonstrate a significant improvement over state-of-the-art fluent speech models.

\subsection{Ablation Study}
	An ablation study was conducted to determine the impact of the multitask architecture and transcript decoder. StutterZero and StutterFormer were re-trained across a five-fold cross-validation with all hyperparameters, data splits, and other tunable values kept constant. However, the transcript decoder was removed from both models during the ablation.
\begin{table*}[ht]
	\centering
	\footnotesize
	\setlength{\tabcolsep}{6pt} 
	\renewcommand{\arraystretch}{1.3} 
	\begin{tabularx}{\textwidth}{X c c c}
		\toprule
		\textbf{Model} & \textbf{WER} & \textbf{CER} & \textbf{Mean BERTScore Precision} \\
		\midrule
		StutterZero (with transcript decoder)   & $0.116 \pm 0.013$ & $0.110 \pm 0.052$ & $0.9174 \pm 0.034$ \\

Ablated StutterZero (without transcript decoder)  & $0.339 \pm 0.011$ & $0.260 \pm 0.012$ & $0.437 \pm 0.010$ \\

StutterFormer (with transcript decoder)   & $0.08 \pm 0.03$ & $0.07 \pm 0.011$ & $0.9411 \pm 0.012$ \\

Ablated StutterFormer (without transcript decoder)  & $0.221 \pm 0.016$ & $0.194 \pm 0.019$ & $0.552 \pm 0.015$ \\
		\bottomrule
	\end{tabularx}
	\caption{Ablation Study Results on Validation Data.}
	\label{tab:ablate}
\end{table*}
	This significant difference in every metric after the ablation in both models demonstrates the importance of the transcript decoder. Observing StutterZero, the WER increased by 22.3\% whilst the CER only increased by 15\%.  This shows the ablated StutterZero predicted most of the characters in a word correctly, but perhaps missed a few characters in some more words. The same phenomenon is seen in StutterFormer ablation, but to a lesser degree. This aligns with the functionality of the transcript decoder -- to help both end-to-end models capture more detailed orthographical features in words. A greater increase in WER than in CER could mean StutterZero/StutterFormer incorrectly predicted more allophones and homophones, which have similar character-level spellings but are different words.

\subsection{FluencyBank Validation Test}
FluencyBank is a subset of the TalkBank corpus, an open repository for spoken language data. FluencyBank specifically focuses on disfluencies, including stuttering. Because the FluencyBank dataset is password-protected and it is not possible to automate the scraping of data from the website, this research manually downloaded audio clips from a randomly selected recording along with their transcripts from the Voice-AdultsWhoStutter (Voices-AWS) subset. After downloading and splitting each audio file to be 30 seconds or less, there were 800 stuttered audio samples.

Because FluencyBank is an entirely new dataset with unique audio characteristics, StutterZero and StutterFormer are tested on data they have never encountered before. This ensures that no bias from the training data influences the results. Table \ref{tab:fbank} shows the WER and BERTScore metrics after testing StutterZero and StuttFormer on samples of the FluencyBank dataset.
\begin{table}[H]
	\centering
	\footnotesize
	\setlength{\tabcolsep}{1.5pt} 
	\renewcommand{\arraystretch}{1.3} 
	\begin{tabular}{l c c c c}
		\toprule
		\textbf{Audio ID} & \textbf{SZ WER} & \textbf{SF WER} & \textbf{SZ BERT} & \textbf{SF BERT} \\
		\midrule
		Mean & $0.161 \pm 0.034$ & $0.120 \pm 0.013$ & $0.915 \pm 0.015$ & $0.937 \pm 0.023$ \\
		\bottomrule
	\end{tabular}
	\caption{Performance comparison of StutterZero (SZ) and StutterFormer (SF) on the FluencyBank dataset using WER and BERTScore Precision metrics.}
	\label{tab:fbank}
\end{table}
	Table \ref{tab:spects} presents a qualitative comparison between the stuttered and fluent spectrograms for three speech samples from the FluencyBank dataset. The regions enclosed by green rectangles highlight sections containing a stutter in the original audio or the corresponding fluent segments after correction.

	The effects of stutter correction are most apparent in the first row (audio 29mb\_1.wav). In the stuttered spectrogram, the doubled upward-sloping signal represents a word repetition (“can you–can you”). Both StutterZero and StutterFormer successfully remove this repetition in their predicted fluent spectrograms, producing a single, continuous signal in place of the duplicated pattern.
\begin{table*}[!t]
\centering
\resizebox{\textwidth}{!}{%
\begin{tabular}{|>{\centering\arraybackslash}m{0.08\textwidth}|
                >{\centering\arraybackslash}m{0.15\textwidth}|
                >{\centering\arraybackslash}m{0.3\textwidth}|
                >{\centering\arraybackslash}m{0.3\textwidth}|
                >{\centering\arraybackslash}m{0.3\textwidth}|}
\hline
\textbf{ID} &
\textbf{Transcript} &
\textbf{Stuttered Spectrogram} &
\textbf{StutterZero Predicted Fluent Spectrogram} &
\textbf{StutterFormer Predicted Fluent Spectrogram} \\ \hline

29mb\_1.wav &
so \textcolor{red}{can you can you} talk about the impact that stuttering has had on your everyday life thinking about things like education your job your family your friends &
\includegraphics[width=0.32\textwidth]{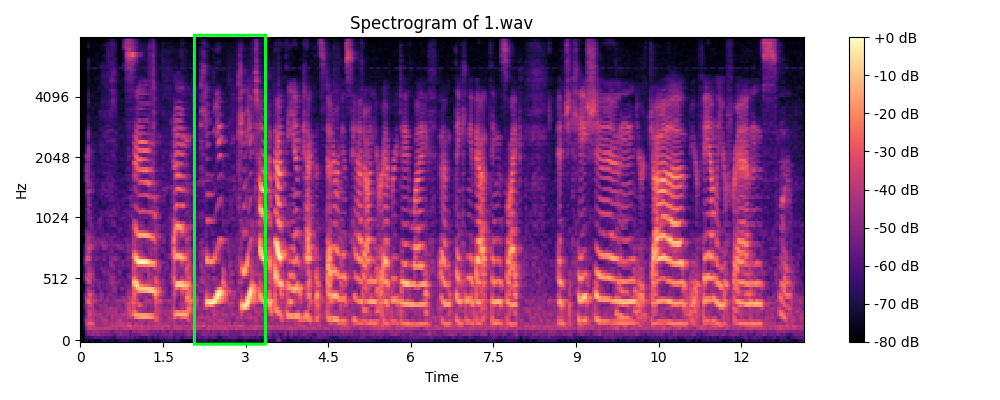} &
\includegraphics[width=0.32\textwidth]{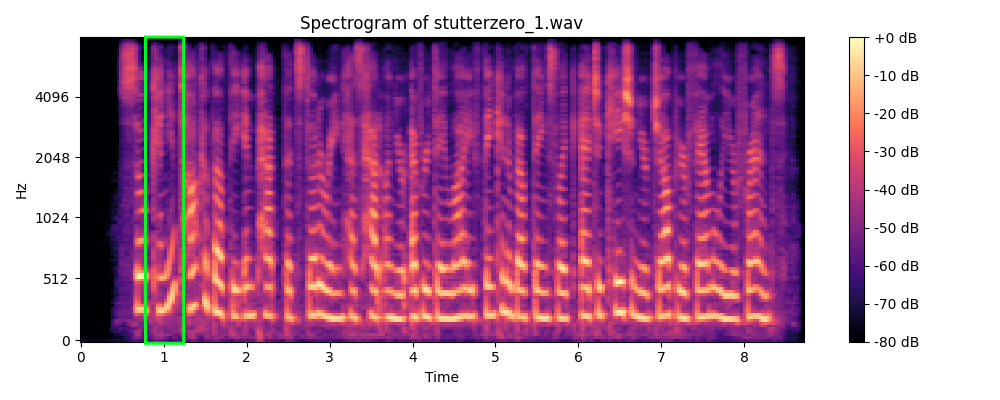} &
\includegraphics[width=0.32\textwidth]{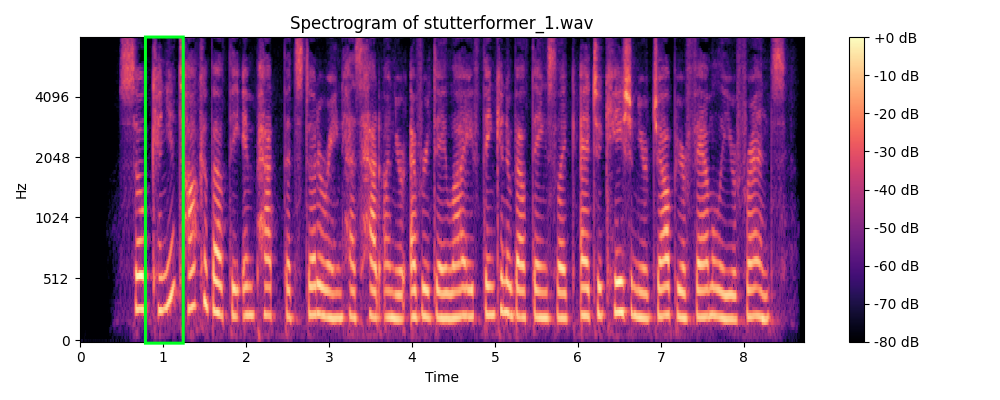} \\ \hline

29mb\_2.wav &
i'd say \textcolor{red}{s-s-s} uhm \textcolor{red}{s-s-s-stuttering} kind of impacted me or like \textcolor{red}{I-I-I-I} had highs and lows &
\includegraphics[width=0.32\textwidth]{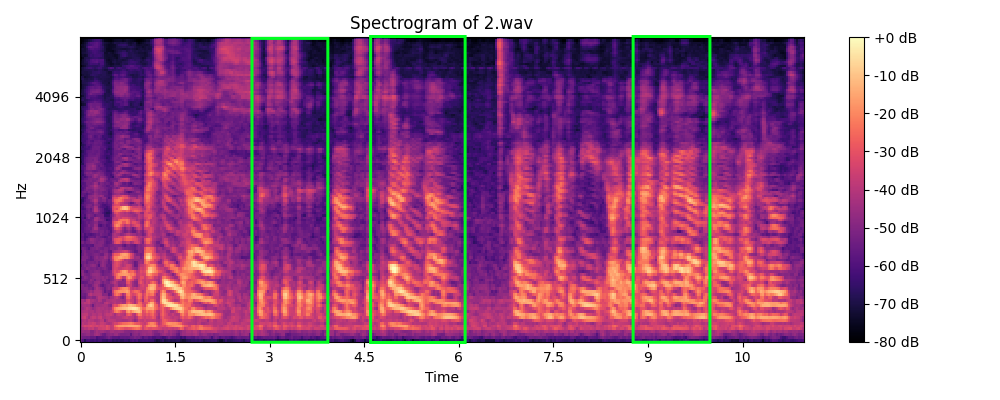} &
\includegraphics[width=0.32\textwidth]{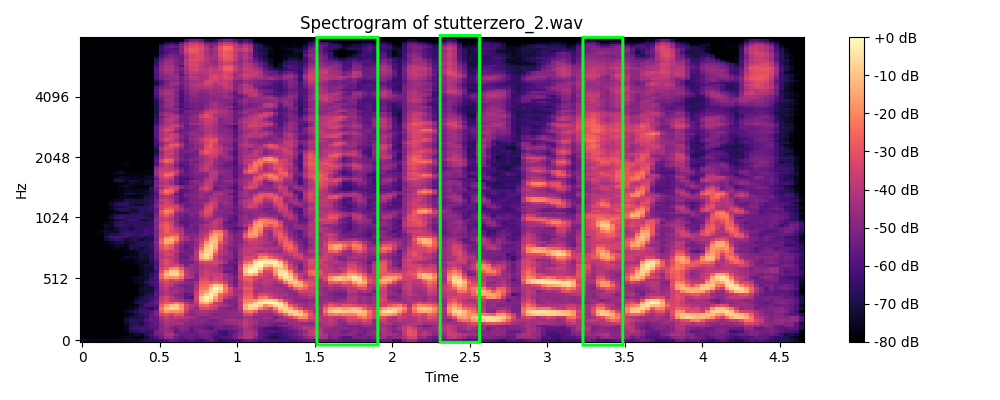} &
\includegraphics[width=0.32\textwidth]{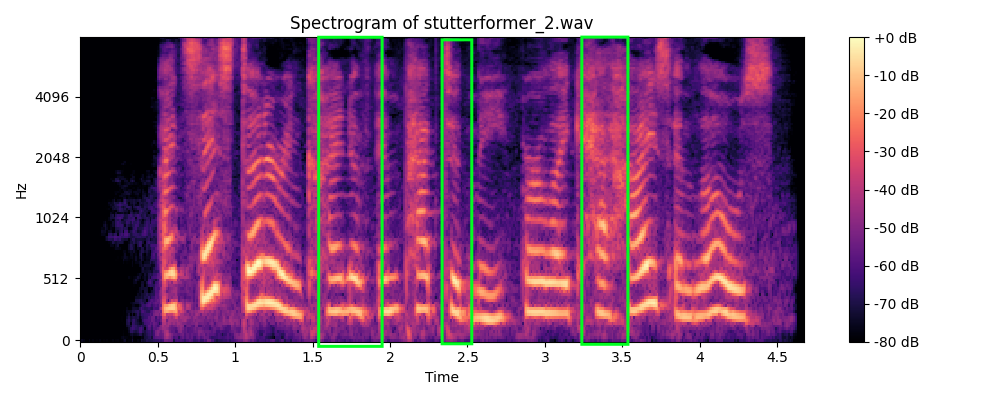} \\ \hline

29mb\_3.wav &
\textcolor{red}{with with} \textcolor{red}{s-s-stuttering} early on \textcolor{red}{d-d-d-definitely} in my elementary school &
\includegraphics[width=0.32\textwidth]{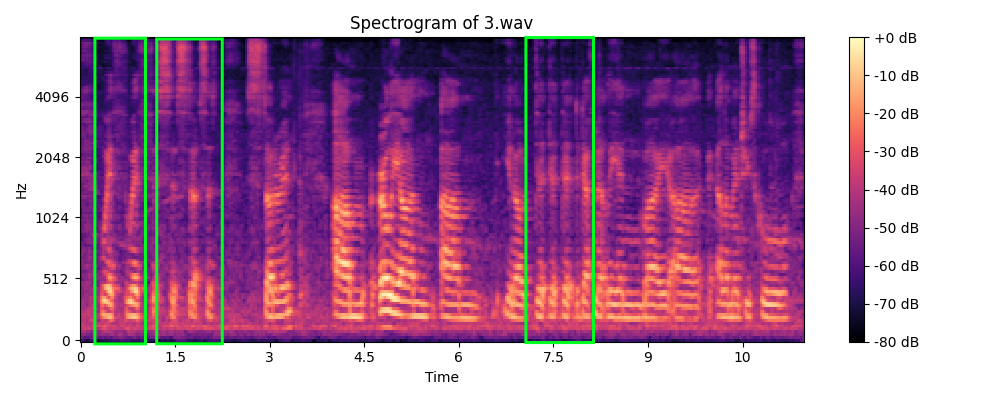} &
\includegraphics[width=0.32\textwidth]{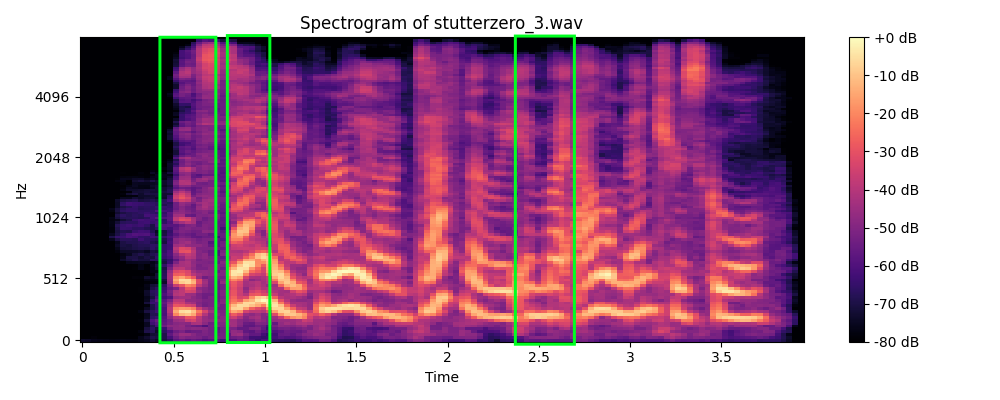} &
\includegraphics[width=0.32\textwidth]{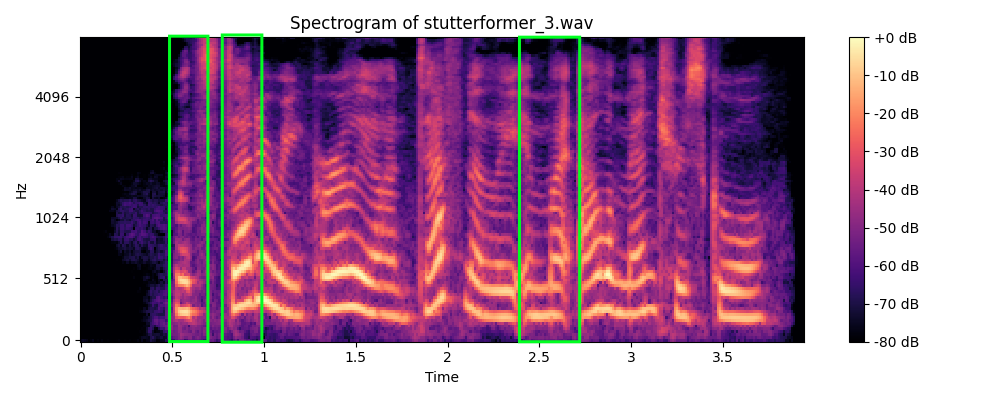} \\ \hline
\end{tabular}%
}
\vspace{2pt} 
\caption{Spectrograms of stuttered audio, StutterZero-predicted fluent audio, and StutterFormer-predicted fluent audio. Areas enclosed by green rectangles indicate either a stutter or the corresponding fluent region after correction.}
\label{tab:spects}
\vspace{-0.8em} 
\end{table*}

\section{Discussion}
This work introduces three stutter correction systems, including the first two end-to-end encoder–decoder models, and provides evidence that direct conversion of stuttered to fluent audio is both feasible and effective.

StutterZero and StutterFormer significantly outperformed state-of-the-art fluent speech recognition models in WER, CER, and BERTScore. The Wilcoxon Signed-Rank Test demonstrated the improvements ($p < 0.05$) of StutterZero and StutterFormer over the next best model, Whisper-Medium. This demonstrates the effectiveness of multitask, end-to-end models for stuttered speech recognition.
	
Unlike existing stutter correction approaches, which struggle to address all five stuttering types, StutterFormer achieves an average transcription accuracy of 90\% on the combined SEP-28K and LibriStutter validation set. It maintains robust performance on entirely new data with different speakers and recording conditions, achieving 88\% accuracy. These results highlight StutterFormer’s resilience to variations in audio quality and speaker prosody, enabling consistently high performance across diverse datasets.
	
StutterZero is not far behind, with an 88\% transcription accuracy on the combined validation set and 84\% on the FluencyBank subset. The slight decrease in accuracy may be attributed to the Transformer architecture’s efficiency in modeling speech and audio sequences. Indeed, correcting stuttered speech is a sequence-to-sequence task, similar to machine translation and other language tasks where Transformers have shown exceptional performance \cite{chorowski2015}.
	
The observed performance improvements resonate with trends in related domains, such as dysarthric and accented speech recognition, where multitask and end-to-end approaches have consistently demonstrated robustness to atypical input \cite{wang2022, mujtaba2025}. By showing that stutter correction benefits strongly from the integration of textual and acoustic objectives, our findings suggest that linguistic supervision is not merely auxiliary but foundational. This echoes psycholinguistic accounts that disfluency cannot be treated as random noise but reflects systematic deviations in speech planning and motor execution. In this sense, our work challenges the assumption -- common in early DSP and ASR pipelines -- that acoustic correction can be divorced from lexical anchoring.
	
The superior performance of StutterFormer over StutterZero cannot be attributed solely to increased model capacity. Rather, Transformer attention provides specific advantages for disfluency correction. Multi-head self-attention captures long-range dependencies, allowing the model to flexibly relate repeated or prolonged segments of speech to their fluent counterparts. This mechanism facilitates alignment across syllables and words, which is essential when disfluencies span multiple phonemic units. In addition, attention layers can simultaneously model both global and local phonetic details, helping to preserve intonation and rhythm while removing interruptions. These properties explain why attention-based architectures are particularly well-suited for mapping stuttered to fluent speech \cite{zeyer2019}.

\subsection{Limitations}
	Several limitations should be noted. There remain potential areas for improvement that can be explored in future research.
\subsubsection{Reliance on TTS-Generated Data}	
A substantial portion of training and evaluation relied on fluent audio generated by TTS systems. While this strategy enabled efficient creation of large-scale paired datasets, it introduced a prosodic mismatch between synthetic and natural speech. As a result, models may have partially learned to adapt to synthetic rhythms and intonation rather than capturing the full variability of natural stuttering. This could limit robustness when applied to spontaneous, emotionally nuanced speech. Future research should mitigate this gap by curating larger collections of natural stutter–fluent pairs, leveraging prosody encoders to capture expressive detail, and employing domain adaptation techniques to reduce distributional bias.
\subsubsection{Need for Increased Dataset Diversity}
	The diversity of the training corpus was constrained by limited access to datasets such as UCLASS or FluencyBank. The absence of these datasets restricted coverage of various speaker demographics, accents, and tones, which in turn limits generalizability. There may be risks of overfitting and undergeneralization when training on only the SEP-28K data from the LibriStutter datasets. Even though five-fold cross-validation and testing on a completely new dataset was used to produce a candid estimate of the true performance both approaches, future steps should focus on data augmentation and training on larger datasets. Additional experiments using diverse corpora such as UCLASS or AS-70 \cite{gong2024} are needed.
\subsubsection{Hardware and Compute Limits}
	The training process was constrained by significant hardware limitations. Since the model was trained on a single GPU with only 10 GB of VRAM, both the batch size and model complexity had to be reduced to make training feasible. This, however, resulted in slower training and unstable convergence of the training loss. With access to more powerful hardware, it would be possible to incorporate more advanced architectural choices -- such as increasing the number of heads in the multi-head attention mechanism -- potentially resulting in improved model accuracy.
\subsection{Future Work}
	While this research achieves impressive preliminary results, it also acts as a proof-of-concept, opening the doors for more advanced end-to-end models in stutter correction.
\subsubsection{Prosody-aware Fine-tuning}
	To alleviate the prosodic loss due to training on TTS-generated fluent speech samples, using multiple encoders to capture the tonal and prosodic content of the stuttered speech could be explored. Multi-encoder models have been explored in dysarthic speech conversion, specifically using a prosody encoder to extract prosodic features \cite{wang2022}. This would enable fluent outputs that retain the speaker’s identity, pitch, and expressive nuance. Additionally,  expanding to multilingual and low-resource languages through cross-lingual pretraining and transfer learning would extend accessibility to a wider global population \cite{gong2024}.
\subsubsection{Dataset Expansion and Multilinguality}
	The current study is limited by access to a small number of corpora. Expanding to diverse datasets such as UCLASS, FluencyBank, and AS-70 \cite{gong2024}, as well as developing multilingual training resources, would enhance model generalizability. Cross-lingual pretraining and transfer learning offer promising strategies to extend accessibility to low-resource and global language communities, ensuring that stutter correction technology benefits a wider population.
\subsubsection{Integration with Clinical Practices}
	StutterZero and StutterFormer have great potential in automating and assisting with delayed auditory feedback (DAF), a common technique used in speech-language therapy. It involves recording and playing back a PWS's speech after a brief delay (usually a few milliseconds to fractions of a  second) \cite{ozker2025} \cite{buzzeti2018}. Playback of the fluent speech can demonstrate how fluent speech is supposed to sound, thereby reinforcing self-monitoring and reducing disfluencies. Instead of the individual hearing their disfluent utterances delayed, the model could provide them with a fluent version of what they intended to say. This would supply the brain with consistent, fluent auditory feedback, potentially reducing the reinforcement of stuttered patterns while strengthening neural pathways associated with fluent production. Indeed, speaking in unison with a fluent signal (an external “fluent version” of one’s speech content) reliably induces near-instant fluency in most PWS \cite{kalinowski2003}.
\subsubsection{Real-Time Communication Applications}
	Beyond therapy, optimized versions of StutterZero and StutterFormer could be deployed in real-time communication settings. With techniques such as model quantization, pruning, and knowledge distillation, lightweight implementations may run on mobile or embedded devices. Potential applications include live correction during phone calls, video conferences, and online meetings, where disfluent speech is automatically converted to fluent audio for listeners. Similar methods could also be applied to voice recording, broadcasting, and sound engineering, eliminating the need for re-recording when disfluencies occur.

\section{Conclusion}
	The research presents several contributions to the field of stutter correction and speech conversion. First, an ASR-based pipeline utilizing fine-tuned Whisper-Small that achieved significant performance improvements, reducing Word Error Rate to 4\% compared to 36.1\% for the baseline Whisper-Medium model. Second, and more importantly, this research introduced StutterZero and StutterFormer, the first two end-to-end stutter correction models that directly convert stuttered speech to fluent audio without intermediate transcription steps. StutterZero was a multitask encoder-decoder architecture using conventional convolution and LSTM layers, reducing Word Error Rate to 11\%. StutterFormer was based on a modern Transformer architecture and reduced Word Error Rate even further to 8\%. Being the first end-to-end models for stutter correction, both StutterZero and StutterFormer pave the way for future development of larger and more accurate end-to-end models. Specifically, the encoder-decoder architecture allows for great flexibility in the choice of encoder and decoder, allowing several multitask encoders and decoders to work in unison to learn different representations of the audio. In industry, this research may pave the way for more accessible human-machine interaction and communication.



\bibliographystyle{IEEEtran}
\bibliography{mybib}

\newpage

\section*{Acknowledgment}
This research originated from personal initiative, as the student researcher has a stutter and sought to address this challenge through computational methods. 
The author expresses sincere gratitude to Mr. Cook, teacher of the Millburn High School Science Research Program, for his invaluable mentorship. Mr. Cook provided detailed guidance on the scientific process — from formulating research questions and establishing goals to conducting literature reviews and writing in accordance with academic standards. Equally important, he encouraged students to reach out to professors and domain experts, a practice that significantly shaped the independent yet rigorous nature of this work.

The author also thanks Professor Chen Zeng of The George Washington University. Professor Zeng generously offered his time in biweekly virtual meetings during the summer, providing feedback on datasets, reviewing model architectures, and suggesting future research directions. His role was purely advisory and uncompensated, and he neither contributed code nor writing. His mentorship strengthened the soundness, depth, and rigor of this work.

Special thanks are also extended to Professor JoAnne Cascia and Dr. Iyad Ghanim of Kean University, specialists in speech-language pathology, who provided critical insights into the clinical relevance and potential applications of this research. Their perspectives grounded the project in real-world practice, highlighting how computational models may complement therapeutic approaches and better serve individuals who stutter. Their contributions were advisory only, without involvement in coding or writing, and were provided entirely without compensation.

The guidance of Mr. Cook, Professor Zeng, Professor Cascia, and Dr. Ghanim reflects generosity and a genuine commitment to supporting student research. The author is deeply grateful for their time, patience, and willingness to share expertise.

Finally, the author acknowledges the broader research community that made this project possible. The datasets employed — including SEP-28k and LibriStutter — are publicly available and de-identified, developed by teams committed to advancing accessibility in speech technology. Without these resources, this project would not have been feasible.







\end{document}